\documentclass[aps,pre,twocolumn,floatfix,amsmath,showpacs]{revtex4-1}
\usepackage{color}
\usepackage{graphicx}
\DeclareGraphicsExtensions{.pdf,.png,.jpg}

\begin{document}

\newcommand{\wq}[1]{\textcolor{blue}{#1}}

\title{
Anomalous diffusion with log-periodic modulation in a selected time interval 
}
\author{L.~Padilla}
\email{lorenapadilla.r@gmail.com}
\author{H.~O.~M\'artin}
\author{J.~L.~Iguain}
\email{iguain@mdp.edu.ar}
\affiliation{Instituto de Investigaciones F\'{\i}sicas de Mar del Plata (IFIMAR) and
Departamento de F\'{\i}sica FCEyN,\\
Universidad Nacional de Mar del Plata, De\'an Funes 3350, 7600 Mar del
Plata, Argentina}

\begin{abstract}

On certain self-similar substrates the 
time behavior of a random walk 
is modulated by logarithmic periodic oscillations on all time scales. 
We show that if disorder
is introduced in a way that self-similarity holds only in average, the 
modulating oscillations are washed out but subdiffusion remains as in the 
perfect self-similar case.  Also, if disorder distribution is appropriately chosen
the oscillations are localized in a selected time interval.
Both the overall random walk exponent and the period of 
the oscillations are analytically obtained and confirmed by Monte Carlo
simulations.

\end{abstract}
\maketitle

The basics of anomalous diffusion on fractal substrata were established some decades ago \cite{ale,ram,ben,hav,bou}. However, 
it has been recently realized that, on some self-similar objects, the time behavior of a
random walk (RW) is modulated by logarithmic-periodic oscillations (see, for example, 
\cite{woess,gra,kro,ace,bab1,bab2,maltz,web,lore2}).  
This kind of modulation is very ubiquitous. Examples 
have been observed in biassed diffusion on 
random systems \cite{bernas,stau0,stau,yu}, earthquake dynamics \cite{huang,saleur}, escape probabilities 
in chaotic maps \cite{pola}, processes on random quenched and fractal media \cite{kut,andra,bab3,saleur2}, diffusion-limited 
aggregates \cite{sor1i}, growth models \cite{huang2}, and stock markets \cite{sor2,van1,van2,van3}.
In spite of this diversity, log-periodicity is considered evidence of a discrete scale invariance \cite{sor1}, which results
from some inherent self-similarity \cite{dou}.

Interestingly, the above described complex behavior can be captured by simple models, as is the case of
 a single particle moving by jumps between nearest-neighbor (NN) 
sites of a one-dimensional lattice. 
Indeed, the particle mean square displacement exhibits a subdiffusive behavior modulated by log-periodic oscillations, 
provided that the hopping rates are distributed in a self-similar way \cite{lore1}.
In this work we introduce disorder, by randomly shuffling those hopping rates, and investigate the
effect of this modification on the RW dynamics.
We give theoretical and numerical evidences that, while subdiffusion survives 
with the same RW exponent as in the ordered substrate, the oscillatory modulation is washed out. 
We also show how the oscillations can be localized in selected time windows by a proper choice of disorder distribution.
We believe that our model incorporates the essential mechanisms for anomalous diffusion modulated by localized oscillations. 
In this sense, it is a minimal model, which can serve for understanding more realistic systems.

\begin{figure}
\includegraphics[width=0.95\linewidth]{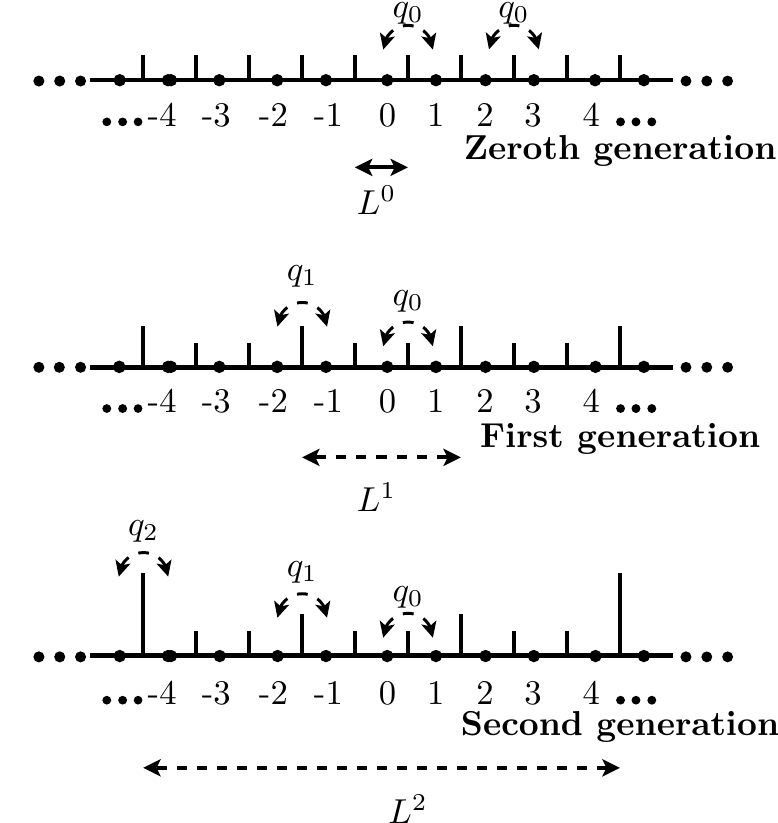}
\caption{Construction of a substrate with a self-similar distribution of hopping rates ($L=3$). 
Top: The zeroth generation ($n=0$) with all the hopping rates equal to $q_0$. Center: 
The first generation ($n=1$), where the hopping rates $q_1$ appear. The corresponding barriers 
are separated by a distance \textit{L}. Bottom: The second generation ($n=2$) shows the emergence 
of the hopping rate $q_2$ separated by a distance $L^2$. In the limit $n \rightarrow \infty$ a full 
self-similar distribution of hopping rates is obtained, which corresponds to a completely ordered 
substrate. Numbers below sites indicate position on the lattice. }\label{reduno}
\end{figure}

{\bf Model - 
}
We consider a RW on a one-dimensional lattice, with transitions between NN 
sites only.
The probability to hop from site $j$ to site $j+1$ (and from site $j+1$ 
to site $j$) per unit time is  denoted by $k^{(j)}$.  
For readability, we schematically represent these rules by a set of barriers of height 
$h^{(j)}=c/k^{(j)}$ ($j\in Z$),  where $c$ is a constant 
(see Fig.~\ref{reduno}).

 {\bf Ordered substrate -}
We briefly review the main properties of the model introduced in Ref.~\cite{lore1}. For more details we refer to that paper.
The substrate depends on two parameters, $L$ (odd integer greater than one) and $\lambda$ (real positive), and
is built in stages. The result of every stage is called a {\it generation} 
(see Fig.\ref{reduno}, for $L=3$).
The zeroth-generation corresponds to the situation in which all the hopping rates 
are identical (\mbox{$k^{(j)}=q_0$,} $\forall j\in Z$). In the first generation, the hopping 
rate $k^{(j)}$ is set to $q_1$ ($<q_0$) for every  $j=pL-(L+1)/2$, with $p$ integer. 
All the other hopping rates remain as in the generation zero.
This iteration continues indefinitely and, in general, the generation $n$ is obtained from the 
generation $n-1$  by setting $k^{(j)}=q_n$ ($<q_{n-1}$), for every 
\mbox{$j=pL^n-(L^n+1)/2$}, with $p$ integer. From now on we
will refer to the full self-similar substrate obtained after an infinite number of iterations
 as a \emph{completely ordered} substrate. In this limit, the dynamics become 
 full self-similar if $D^{(n)}/{D^{(n+1)}}=1+\lambda$ ($n=0,1,2,...$),  
where $D^{(n)}$ is the diffusion constant of the $n$th generation.
This condition is equivalent to

\begin{equation}
\frac{q_{0}}{q_{i}}=\frac{q_{0}}{q_{i-1}}+(1+\lambda)^{i-1}\lambda L^{i},\;\;\;\mbox{for}\; i=1,2,3,...,
\end{equation}
which determine the hopping rates for the completely ordered substrate.

\begin{figure}
\includegraphics[width=0.95\linewidth]{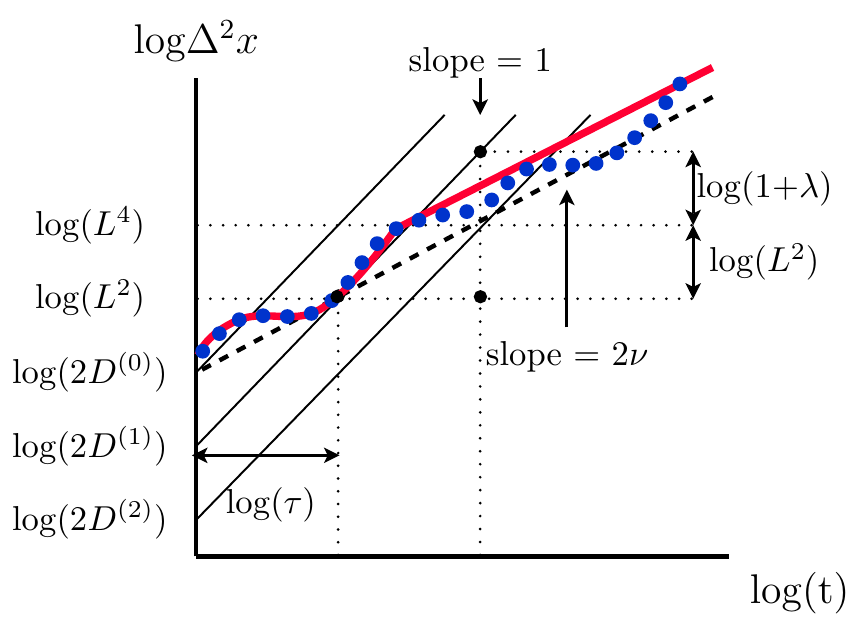}
\caption{(Color online) Mean-square displacement as a function of time. The full straight lines have  slopes of $1$, 
and represent normal diffusion, i. e., $\Delta^2x=2D^{(n)}t$. The dashed  
line represents the global behavior $\Delta^2x \sim t^{2\nu}$.
A log-periodic modulation is observed on all time scales for a completely ordered substrate (blue dots) and
at short times for a partially ordered one (red thick full line).} 
\label{osci}
\end{figure}

It was also demonstrated that, if ${\Delta^2 x(t)}$ is the RW mean-square displacement on the completely ordered substrate,
for a time such that 
$A L^n < (\Delta^2 x(t))^{1/2} < A L^{n+1}$ ($A$ is a constant of the order of one), 
everything will happen as on the $n$th-generation substrate and 
$
\Delta^2 x(t) =  2 D^{(n)} t,
$
for $t$ in that interval. Moreover, because of full self-similarity, the sequence of
length-dependent diffusion coefficients leads to 
\begin{equation}
\begin{centering}
 \Delta^2x (t) = K t^{2 \nu} f(t) {\rm ,}
\end{centering} \label{function}
\end{equation}
where $K$ is a constant, $\nu=[2+\log(1+\lambda)/\log L)]^{-1}$ is the RW exponent,
and $f(t)$ is a log-periodic function of  period 
$\tau=L^{1/\nu}$ (i.~e., $f(t\tau)=f(t)$).
This behavior is shown qualitatively in Fig.~\ref{osci} (dots).

\begin{figure}
\includegraphics[width=0.95\linewidth]{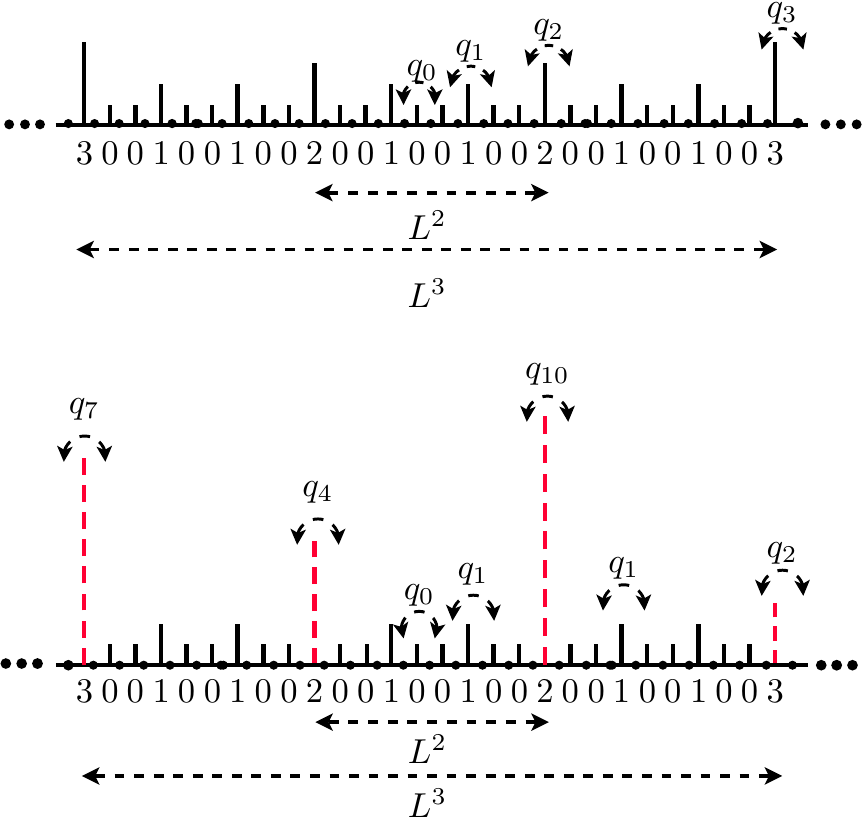}
\caption{(Color online) Top:
Schematic of a completely ordered substrate ($L=3$). Numbers bellow barriers
indicate NN pair label.  
Bottom: A partial disorder realization of the substrate in the top ($p=0$ and $m=1$). The locations of 
$q_0$ and $q_1$ (black full lines) remain fixed. The others hopping rates (red-dashed lines) are 
distributed at random on the rest of the lattice.}
\label{od}
\end{figure}

{\bf Disorder -}
For the sake of clarity, when $q_l$ is the hopping rate between sites $j$ 
and $j+1$ in the perfectly ordered substrate, we will label pair of sites
 with the number $l$.
Note that the set of NN pairs labeled with $l$ 
can be roughly associated with the length scale $L^{l}$ and that $q_l$ appears on the lattice with 
a frequency $f_l=(L^{l+1})/(L-1)$ (see  Fig.~\ref{od}-top).

A disorder realization will implicate a random permutation of the set $\{q_i | i\ge 0\}$, on the same one-dimensional
lattice. In this context, we will study the disorder average of the RW mean-square displacement. 
It is useful to distinguish two different scenarios. When the entire array of hopping rates is
randomized on the lattice we will refer to the resulting substrate as
being \emph{completely disordered}. In contrast, we will say that it is \emph{partially ordered}
if only the hopping rates in a subset of the ordered array are randomly permuted.

In each realization of the completely disordered substrate,
the hopping rate between any pair of NN sites is 
$q_i$ with a probability $P_i=f_i$ (for $i\ge 0$), consistent with the frequency distribution. 
It is not hard to see that the log-periodic modulation, present
in the case of the self-similar substrate, should be washed out because of disorder.
For the self-similar case, log-periodicity comes from the
 strong association between
the diffusion constant $D^{(n)}$ and length scales between $L^n$ and $L^{n+1}$ (see Fig.~\ref{osci}),
and, thus, should disappear when the special order in the array of hopping rates is lost.
On the other hand, since we are always dealing with the same set $\{q_i | i\ge 0\}$, 
the average spreading speed (i.~e., the speed at which a RW  explores the lattice) can not be
affected by disorder \cite{ben}. Therefore, we expect that the disorder-averaged mean-square displacement 
$ \overline{\Delta^2x} $ behaves, as a function of time, as

\begin{equation}
\begin{centering}
 \overline{ \Delta^2x}(t) = E t^{2 \nu}  {\rm ,}
\end{centering} \label{global}
\end{equation}
($E$ is a constant) with the same RW exponent  $\nu$ as in the perfectly ordered substrate but without oscillations.

A partially ordered (for length scales between $L^p$ and $L^{p+m+1}$, with $m \geq 1$) substrate
can be defined as follows. The hopping rate for every NN pair with its label in  $\{p, p+1,...,p+m\}$ is the same as in the completely ordered case. 
That is, for any pair labeled with
$l$ in that set, the corresponding hopping rate is $q_l$. For the other 
NN pairs,  $q_i$ (with $i \notin \{p, p+1,...,p+m\}$) is randomly chosen  from the set \{$q_i, with  i \notin \{p, p+1,...,p+m$\}\}
with probability $P_i= f_i/Z$,
where $Z=\sum_{i \notin \{p, p+1,...,p+m\}}f_i$.
Thus, in
every realization of disorder,
the  hopping rates associated with length scales between $L^p$ 
and $L^{p+m}$ remain as in the completely ordered substrate but the other
are randomly distributed on the rest of the lattice.

According to the discussion above we expect that disorder-averaged mean-square displacement will continue 
to show an overall subdiffusive behavior, with the same RW exponent as in the completely ordered substrate and 
that, due to partial order,  an oscillatory modulation appears  
when $B L^p < ({\overline{\Delta^2 x}(t)})^{1/2} < B L^{p+m+1}$,   
(where $B$ is a constant). Outside this interval, the oscillations should be erased by disorder.  
Note that
the geometrical analysis leading to the relation among $\nu$, $L$ and $\tau$ \cite{lore1} can be performed at any scale
 and, thus, $\tau=L^{1/\nu}$, also in this case.

To summarize, we predict a subdiffusive behavior modulated by oscillations in a time window determined by disorder
distribution.
For concreteness, let us assume $p=0$. In this case, the initial ordered array 
remains 
without changes up to the length scale $L^{m+1}$. On larger scales, 
the hopping rates are randomly distributed.
(see the example with $L=3$ and $m=1$ shown in Fig.~\ref{od}).
We thus  expect the behavior sketched in  
Fig.~\ref{osci}. That is, anomalous diffusion (with $2 \nu <1$), modulated by oscillations at short times 
and without oscillations at longer times.

\begin{figure}
  \includegraphics[width=0.95\linewidth]{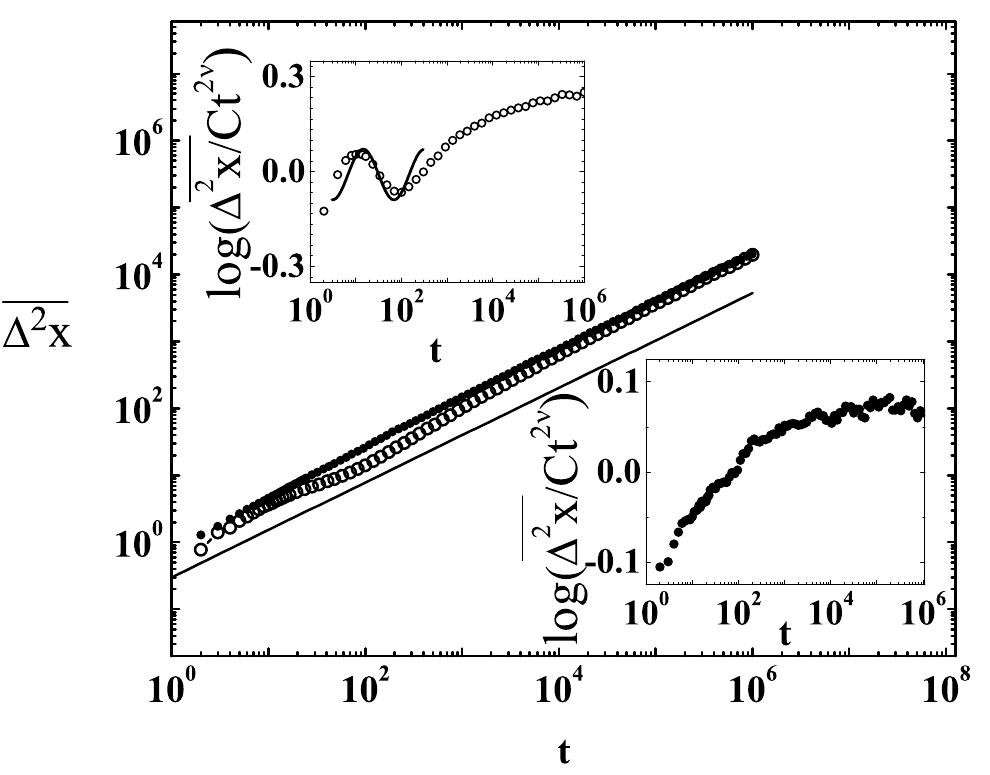}\\
  \caption{$\overline{\Delta^2x}$ against $t$, for a 
completely disordered substrate (filled circles) and for partially ordered one with $p=0$ and $m=1$ 
(open circles). In both cases, $L=3$, and $\lambda=1.48$. The straight line has the slope $2\nu$, which
represents the overall subdifussive behavior.
In the insets we show the scaled disorder-averaged mean-square displacement 
as a function of time for each case (symbols as in the main panel). The curve in the upper inset
corresponds to a harmonic function with period $\tau$. Both $\nu=0.354$ and $\tau=22.3$ were analytically obtained.
For more details, see the text.}\label{c2}
\end{figure}

{\bf Numerical results -}
To test the predictions outlined above, 
we perform standard MC simulations on a lattice of linear size $L^{n}$
($q_0=1/2$ and the time step $\delta t=1$).
Every RW starts at the center of the lattice and $n$ is always large enough 
to prevent the RW from reaching the lattice borders.

The disorder-averaged mean-square displacement as a function of time, obtained for 
structures with  $L=3$ and $\lambda=1.48$, is plotted in Fig.\ref{c2}. The filled circles correspond
to complete disorder and the open circles to partial order with
$p=0$ and $m=1$. 
The straight line has a slope $2\nu$ (with the theoretical value of $\nu$) and was drawn to guide the eyes.
The good agreement between analytical and numerical results is apparent in both cases.
Let us remark that, for the  partially ordered substrate, $\overline{\Delta^2x}(t)$
yields a modulated power-law behavior at short times. 
At longer times, the modulation is erased; however, anomalous diffusion 
still holds,
with the same RW exponent.  In the insets, we have plotted 
$\log [\overline{\Delta^2x} /(Ct^{2\nu})]$ against $\log (t)$, using the same data. 
The constant \textit{C} was appropriately chosen 
to have, in the case of partial order, the oscillations centered around zero. 
In the upper inset, we have 
also drawn the curve $A \sin(2\pi \log(t)/\log(\tau)+\alpha)$, with the
theoretical value of $\tau$. It is a first-harmonic approximation of a $\log(\tau)-$periodic function, consistent with the numerical findings (\textit{A} and $\alpha$ are fitted parameters).

\begin{figure}
\includegraphics[width=0.95\linewidth]{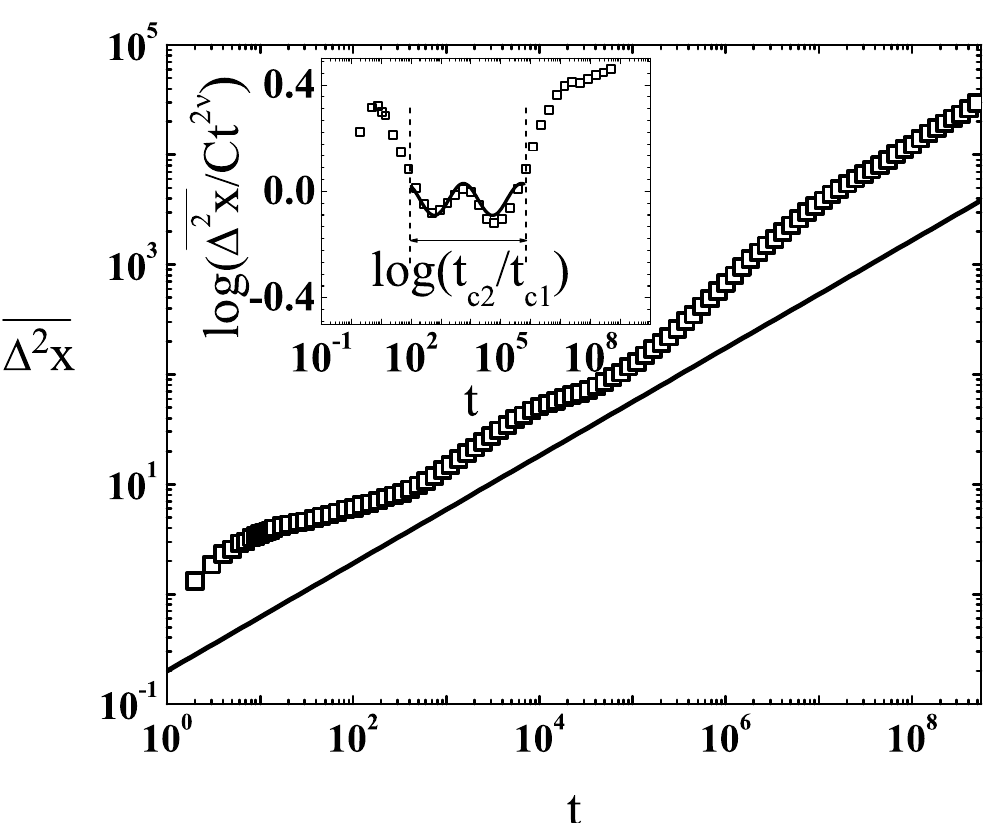}\\
\caption{The disorder-averaged mean-square displacement as a function of time for 
 $L=3$, $\lambda=9$, $p=2$ and $m=1$.  The straight line  has a slope $2\nu$. 
Inset: $\overline{\Delta^2 x}/(C t^{2\nu})$ against $t$ for the same data. 
The curve was drawn as in Fig.~\ref{c2}. Here,  we use  $t_{c2}/t_{c1}=L^{2/\nu}$.
The RW exponent and the period have the analytical values $\nu=0.245$ and $\tau=90.0$.}\label{medio}
 \end{figure}
 
We have observed that if the randomness does not affect the structure 
on length scales smaller than a
cutoff distance $l_c$,  the oscillations are present  only for times shorter than a crossover time
$t_c$ ($\sim l_c^{1/\nu}$). To get a modulation in another time window we 
explore diffusion a substrate with $p=2$, $m=1$, $L=3$, and $\lambda =9$.  
The corresponding numerical results, plotted in Fig.\ref{medio}, show the
oscillating amplitude localized at the center of the graph; 
outside this region the time dependence of $\overline{\Delta^2x}$ is well described by a power law. As in previous figures, 
the inset shows a filtered plot and a first-harmonic approximation of the data. The modulation is clearly
localized between two crossover times $t_{c1}$ and $t_{c2}$, with $t_{c2}/t_{c1}\simeq L^{(1+m)/\nu}$. Note that,  also in this
case, all our predictions are consistent with the numerical findings (Fig.~\ref{medio}-inset).

{\bf Conclusions - 
}
When disorder is introduced into a self-similar structure, a single particle
diffusing on the resulting substrate exhibits an anomalous 
behavior modulated by oscillations in some characteristic time intervals.
Both the RW exponent and the period of the oscillations are determined by the parameters of the original perfectly ordered substrate. On the other hand, the position and length of the time window in which the oscillatory behavior survives randomization 
depend on disorder distribution. Indeed, the modulation holds for those length scales that are not affected by disorder; however,
it is washed out on other scales because of the incoherent superposition involved in the averaging process.
Finally, we would like to emphasize that even though we have focused on oscillations located in a continuous interval, the generalization to the case of several time windows separated by gaps of unmodulated behavior is straightforward.
Thus, the oscillatory regime can be modified as desired by proper choice of $L$, $\lambda$ and disorder distribution.

{\bf Acknowledgments -}
This work was supported by the Universidad Nacional de Mar del Plata and the 
Consejo Nacional de Investigaciones Cient\'{\i}ficas y T\'ecnicas --CONICET-- 
(PIP 0041/2010-2012).

\end{document}